\documentclass[letterpaper, 10 pt, conference]{ieeeconf}  

\IEEEoverridecommandlockouts                              

\usepackage{graphics} 
\usepackage{epsfig} 
\usepackage{amsmath} 
\usepackage{comment}

\title{\LARGE \bf
A comparison of methods to eliminate regularization weight tuning from data-enabled predictive control
}

\author{Manuel Koch$^{1}$ and Colin N. Jones$^{2}$
\thanks{$^{1}$Manuel Koch is with the Department of Electrical Engineering, EPFL, 1015 Lausanne, Switzerland and with the Faculty of Architecture, Civil Engineering and Geomatics, FHNW, 4132 Muttenz, Switzerland
        {\tt\small manuelpascal.koch@epfl.ch/manuel.koch@fhnw.ch}}%
\thanks{$^{2}$Colin Jones is with the Department of Electrical Engineering, EPFL, 1015 Lausanne, Switzerland
        {\tt\small colin.jones@epfl.ch}}%
}

\usepackage[dvipsnames]{xcolor}

\begin{document}

\maketitle
\thispagestyle{empty}
\pagestyle{empty}

\begin{abstract}

Data-enabled predictive control (DeePC) is a recently established form of Model Predictive Control (MPC), based on behavioral systems theory. While eliminating the need to explicitly identify a model, it requires an additional regularization with a corresponding weight to function well with noisy data. The tuning of this weight is non-trivial and has a significant impact on performance. In this paper, we compare three reformulations of DeePC that either eliminate the regularization, or simplify the tuning to a trivial point. A building simulation study shows a comparable performance for all three reformulations of DeePC. However, a conventional MPC with a black-box model slightly outperforms them, while solving much faster, and yielding smoother optimal trajectories. Two of the DeePC variants also show sensitivity to an unobserved biased input noise, which is not present in the conventional MPC. 

\end{abstract}

\section{INTRODUCTION}

Data-enabled predictive control (DeePC) is a reformulation of model predictive control (MPC), based on behavioral systems theory that was first proposed by Coulson in 2019 \cite{coulsonDataEnabledPredictiveControl2019}. In classical black-box MPC, a model of the controlled plant is identified from operational data, which is then used in the optimization of a control target. DeePC effectively merges both steps, synthesizing an optimized input-output trajectory directly as a linear combination of measured system trajectories. One of the challenges with the original formulation is the selection of a regularization weight, which has been shown to have a significant impact on the performance \cite{coulsonDataEnabledPredictiveControl2019, dorflerBridgingDirectIndirect2022, berberichDatadrivenModelPredictive2021, kochComparisonBehavioralSystems2023}. A literature search has yielded three published reformulations of the basic DeePC to avoid this problem, using an alternative regularization \cite{dorflerBridgingDirectIndirect2022}, a bi-level formulation of the optimization problem \cite{lianAdaptiveRobustDatadriven2021}, and an extension with an instrumental variable \cite{vanwingerdenDataenabledPredictiveControl2022} respectively. In this study, we compare these methods to an MPC based on an autoregressive model with exogenous inputs (ARX) in a simulated building temperature tracking task, and discuss them from a practical application point of view. 

\section{METHODOLOGY}

\cite{coulsonDataEnabledPredictiveControl2019} formulates a predictive controller based on Willems' fundamental lemma \cite{willemsNotePersistencyExcitation2005} as follows: An initialization length $T_{ini}$ and a prediction horizon $T_f$ are chosen. Given the input and output trajectories $u_{1:T}$ and $y_{1:T}$ with $T$ elements each, the Hankel matrices (\ref{eq:Hu}) and (\ref{eq:Hy}) of depth $L = T_{ini} + T_f$ and width $T - L + 1$ are built. The data is persistently excited of order $T_{ini}+T_f$ if (\ref{eq:Hu}) has full row rank. The Hankel matrices are then split horizontally into a ``past'' component of depth $T_{ini}$, denoted $p$, and a ``future'' component of depth $T_f$, denoted $f$. 

\begin{equation} \label{eq:Hu}
   \mathcal{H}_L[u_{1:T}] = \begin{bmatrix} U_p \\ U_f \end{bmatrix} = 
       \begin{bmatrix} u_1 & u_2 & \cdots & u_{T-L+1} \\
                       u_2 & u_3 & \cdots & u_{T-L} \\
                       \vdots & \vdots & \ddots & \vdots \\
                       u_L & u_{L+1} & \cdots & u_T \end{bmatrix}
\end{equation}

\begin{equation} \label{eq:Hy}
    \mathcal{H}_L[y_{1:T}] = \begin{bmatrix} Y_p \\ Y_f \end{bmatrix} 
\end{equation}

Further, the initialization vectors $u_{ini}$ and $y_{ini}$, each consisting of $T_{ini}$ samples of the input and output trajectories, and the optimization variables $u$ and $y$ of length $T_f$ are introduced. These are combined into the basic DeePC formulation (\ref{eq:deepc}) with the additional optimization variable $g$ with $T - L + 1$ elements, and a quadratic regularizer with the tuning weight $\lambda$. We abbreviate this into the combined Hankel matrix $\mathcal{H}$ and the right-hand side vector $v$. For legibility, we use a generic cost function $J(u,y)$. In the following sections, $\dagger$ denotes the Moore-Penrose inverse. 

\begin{align}
    & \min_{u,y,g} J(u,y) + \lambda \lVert g \rVert_2  \label{eq:deepc}  \\
    & s.t. \nonumber \\
    & \mathcal{H} g = 
      \begin{bmatrix} U_p \\ Y_p \\ U_f \\ Y_f \end{bmatrix} g = 
      \begin{bmatrix} u_{ini} \\ y_{ini} \\ u \\ y \end{bmatrix} = v \nonumber \\
    & u \in \mathcal{U}, y \in \mathcal{Y} \nonumber 
\end{align}

\subsection{Extension to disturbance with forecast}

For the following sections, we introduce the disturbance variable $w$ with the Hankel matrix $W$ split into $W_p$ and $W_f$, as well as the disturbance initialization $w_{ini}$ and forecast $w_f$. $\mathcal{H}$ and $v$ from equation (\ref{eq:deepc}) are thus extended to (\ref{eq:extH}) and (\ref{eq:extV}). We further define $\hat{\mathcal{H}}$ the first five elements of $\mathcal{H}$, and $\hat{v}$ the first five elements of $v$ of the extended formulations. 

\begin{equation} \label{eq:extH}
    \mathcal{H} = \begin{bmatrix} \hat{\mathcal{H}} \\ Y_f \end{bmatrix} = 
    \begin{bmatrix} U_p^\top \hspace{-2mm} & W_p^\top \hspace{-2mm} & Y_p^\top \hspace{-2mm} & U_f^\top \hspace{-2mm} & W_f^\top \hspace{-2mm} & Y_f^\top \end{bmatrix}^\top
\end{equation}

\begin{equation} \label{eq:extV}
    v = \begin{bmatrix} \hat{v} \\ y \end{bmatrix} = 
    \begin{bmatrix} u_{ini}^\top \hspace{-2mm} & w_{ini}^\top \hspace{-2mm} & y_{ini}^\top \hspace{-2mm} & u^\top \hspace{-2mm} & w_f^\top \hspace{-2mm} & y^\top \end{bmatrix}^\top
\end{equation}

\subsection{Method 1: Projection to least-squares identification}

The method described in \cite{dorflerBridgingDirectIndirect2022} replaces the quadratic regularization with an orthogonal projector on a least-squares identification, i.e. punishing a deviation therefrom while not restricting degrees of freedom in line with it. The regularization weight can thus be chosen very high without affecting the control optimization, effectively eliminating the tuning task. The original publication also contains a numerical study of the method on a benchmark single-input, single-output, 5th order, linear time-invariant system. 

With the orthogonal projector $\Pi = \mathcal{H}_1^\dagger \mathcal{H}_1$ and $(I - \Pi)$: 

\begin{align}
    & \min_{u, y, g} J(u,y) + \lambda_g \lVert (I - \Pi) g \rVert  \\
    & s.t. \nonumber \\
    & \mathcal{H} g = v \nonumber \\
    & u \in \mathcal{U}, y \in \mathcal{Y} \nonumber 
\end{align}

\subsection{Method 2: Bi-level optimization}

A robust bi-level formulation of DeePC is described in \cite{lianAdaptiveRobustDatadriven2021}. The setup consists of an inner optimization, which essentially identifies a multi-step model, and an outer optimization, which optimizes a control trajectory. Under a convex reformulation, this is solved as the single-level optimization problem (\ref{eq:bilevel} - \ref{eq:Kmatrix}).  $\tilde{w}$ is the nominal disturbance forecast and $\tilde{\mathcal{W}}$ the corresponding uncertainty. The diagonal form of $K$ enforces causality. $\kappa$ is a dual variable and $\varepsilon_g$ a relaxation of the output noise explained in detail in the original publication, which also reports the successful application of the method to building control in both simulation and experimentation. $\Sigma_v$ the variance of the measurement noise, and $I_{n_y}$ the identity matrix of the system output size $n_y$. 

\begin{align}
    & \min_{\bar{u}, y, K} J(u, y) \label{eq:bilevel} \\
    & s.t. \nonumber \\
    & \tilde{w}_f \in \bar{w} \oplus \tilde{\mathcal{W}} \nonumber \\
    & u = \bar{u} + K \tilde{w}_f \in \tilde{\mathcal{U}} \nonumber \\
    & y = Y_f g(\tilde{w}_f) \in \tilde{\mathcal{Y}} \nonumber \\
    & \begin{bmatrix} g(\tilde{w}_f) \\ \kappa \end{bmatrix}
      = M^{-1} \begin{bmatrix} Y_p ^\top y_{ini} \\ u_{ini} \\ \tilde{w}_{ini} \\ u \\ \tilde{w}_f \end{bmatrix} \nonumber \\
    & (Y_p^\top Y_p + \varepsilon_g) g + H^\top \kappa - Y_p^\top y_{ini} = 0 \nonumber 
\end{align}

with

\begin{equation} \label{eq:bleps}
    \varepsilon_g = T_{ini}^2 tr(\Sigma_v) I_{n_y}
\end{equation}

\begin{equation}
    H := \begin{bmatrix} U_p^\top \hspace{-2mm} & W_p^\top \hspace{-2mm} & U_f^\top \hspace{-2mm} & W_f^\top \end{bmatrix}^\top
\end{equation}

\begin{equation}
    M := \begin{bmatrix} Y_p^\top Y_p + \varepsilon_g & H^\top \\
         H & 0 \end{bmatrix}
\end{equation}

\begin{equation} \label{eq:Kmatrix}
    K =\begin{bmatrix}
    0 & 0 & 0 & \cdots & \vdots \\
    K_{2,1} & 0 & 0 & \cdots & \vdots \\
    K_{3,1} & K_{3,1} & 0 & \cdots & \vdots \\
    \vdots & \ddots & \ddots & \ddots & \vdots \\
    K_{n_h,1} & K_{n_h,2} & K_{n_h,3} & \cdots & 0 
    \end{bmatrix}
\end{equation}

In contrast to the original publication, we add a small quadratic cost with a weight of $10^{-10}$ on all elements of the optimization variables $g$, $\kappa$, $\bar{u}$ and $K$. Without this modification, the optimization occasionally fails in our simulation. 

\subsection{Method 3: Instrumental variable extension}

The instrumental variable formulation of DeePC in \cite{vanwingerdenDataenabledPredictiveControl2022} projects the optimization problem into fewer dimensions. The reduced optimization variable $\hat{g}$ has no more physically superfluous degrees of freedom, thus eliminating the need for regularization. As we show below, the original formulation (\ref{eq:ivorig}) can be reshaped to eliminate $\hat{g}$ altogether. The original publication reports a simulation study using a 5th order linear, time-invariant system describing a simple mechanical test setup. 

\begin{align}
    & \min_{u, y, \hat{g}} J(u, y) \label{eq:ivorig} \\
    & s.t. \nonumber \\
    & \hat{\mathcal{H}} \hat{\mathcal{H}}^\top \hat{g} = \hat{v} \nonumber \\
    & Y_f \hat{\mathcal{H}}^\top \hat{g} = y \nonumber \\
    & u \in \mathcal{U}, y \in \mathcal{Y} \nonumber 
\end{align}

The equality constraints can be reformulated to the non-causal, linear multi-step predictor:

\begin{equation}
    y = f(u) = Y_f \hat{\mathcal{H}}^\top (\hat{\mathcal{H}} \hat{\mathcal{H}}^\top)^{-1} \hat{v} = Y_f \hat{\mathcal{H}}^\dagger \hat{v}
\end{equation}

The MPC problem thus becomes:

\begin{subequations}
\begin{align}
    & \min_{u, y} J(u, y)  \\
    & \text{s.t.} \nonumber \\
    & y = Y_f \hat{\mathcal{H}}^\dagger \hat{v} \label{eq:iv2b}\\
    & u \in \mathcal{U}, y \in \mathcal{Y}  
\end{align}
\end{subequations}

\subsection{Reference method}

An MPC with an ARX model of the same initialization length as the DeePCs is chosen as a well-established reference method. This is an input-output model without hidden states, which makes for a good comparison to DeePC. The model is identified using the corresponding function in Matlab \cite{EstimateParametersARX} with default settings. This results in the reference MPC (\ref{eq:refmpc}) with the ARX model written as $y_{k+1} = f(...)$ for legibility. 

\begin{align}
    & \min_{u,y} J(u,y) \label{eq:refmpc} \\
    & s.t. \nonumber \\
    & u_{ini} = u_{[k-T_{ini}+1:k-1]} \nonumber \\
    & w_{ini} = w_{[k-T_{ini}+1:k]} \nonumber \\
    & y_{ini} = y_{[k-T_{ini}+1:k]} \nonumber \\
    & y_{k+1} = f(u_{ini}, w_{ini}, y_{ini}, u_k) \nonumber \\
    & u \in \mathcal{U}, y \in \mathcal{Y}  \nonumber \\
    & \forall k \in \{ 0, ..., T_f - 1 \} \nonumber 
\end{align}

\subsubsection{Comparison with instrumental variable extension}

The definition of an ARX model from chapter 4.1 of \cite{soderstrom1989system} with the output data $y(t)$, the input data $\varphi(t)$, and the parameter vector $\theta$: 

\begin{equation} \label{eq:arx1}
    y(t) = \varphi(t) \theta \\
\end{equation}

With identification data written in matrix form as:

\begin{equation}
    Y = \Phi \theta
\end{equation}

$\Phi$ is a Hankel matrix of the input data, $Y$ is a vector of the output data. Solving for $\theta$ and plugging into (\ref{eq:arx1}) yields:

\begin{equation} \label{eq:arxBook}
    y(t) = \varphi(t) \Phi^\dagger Y 
\end{equation}

The definitions result in a reversed order of the variables, but the similarities between equations (\ref{eq:arxBook}) and (\ref{eq:iv2b}) are apparent: The single-step input vector $\varphi(t)$ corresponds to the multi-step input vector $\hat{v}$, the Hankel matrix of the identification data $\Phi$ to $\hat{\mathcal{H}}$, the vector of the identification data $Y$ to the Hankel matrix $Y_f$, and the single-step output $y(t)$ to the multi-step output $y$. Hence, the instrumental variable DeePC can be thought of as a behavioral approach equivalent to a multi-step ARX. 

\section{BUILDING SIMULATION}

\begin{figure}[thpb]
  \centering
  \includegraphics[width=3in]{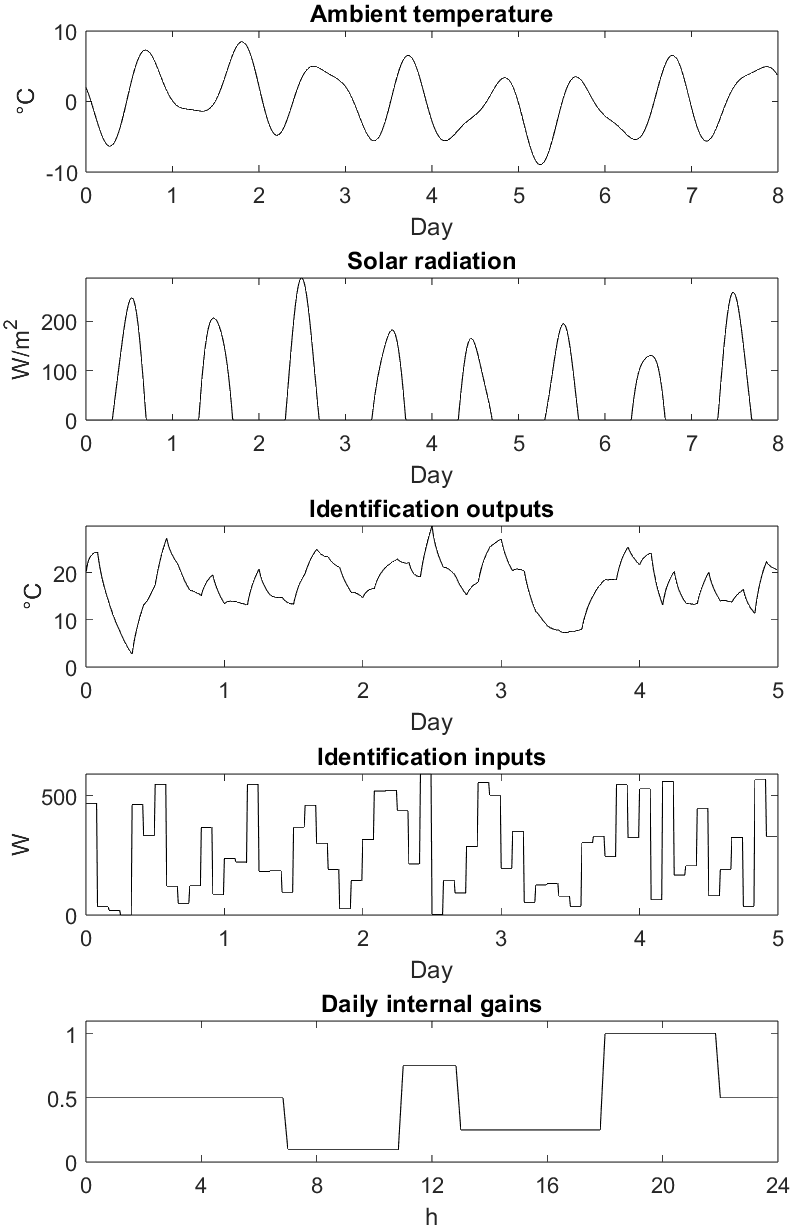}
  \caption{Inputs and outputs for the system identification, respectively construction of the Hankel matrices}
  \label{fig:wea}
\end{figure}

A simplified single-zone building model consisting of (\ref{eq:buildEq}) to (\ref{eq:buildC}) is taken from \cite{oldewurtelTractableApproximationChance2008}. It is of 3rd order with one heating/cooling input, three disturbances (ambient temperature, solar radiation and internal gains), and the zone temperature as observable output. The other two states correspond do an internal and an external wall. The time step is $10 min$. The inputs and disturbances for this study are tuned to yield reasonable values for cold, moderately sunny days with low internal gains (heat from people and devices), explicitly shown in Fig. \ref{fig:wea}. The model appears to be very sensitive to solar gains. The internal gains plot has no unit since it is unclear from the original publication. 

\begin{align} 
    & x_{k+1} = A x_k + B u_k + E w_k \label{eq:buildEq} \\
    & y_k = C x_k
\end{align}

\begin{equation}
    A = \begin{bmatrix} 0.8511 & 0.0541 & 0.0707 \\
                        0.1293 & 0.8635 & 0.0055 \\
                        0.0989 & 0.0032 & 0.7541 \end{bmatrix}
\end{equation}

\begin{equation}
    B = \begin{bmatrix} 0.0035 & 0.0003 & 0.0002 \end{bmatrix}^\top
\end{equation}

\begin{equation}
    E = 10^{-3} \begin{bmatrix}  22.2170 & 1.7912 &  42.2123 \\
                                  1.5376 & 0.6944 &   2.9214 \\
                                103.1813 & 0.1032 & 196.0444 \end{bmatrix}
\end{equation}

\begin{equation} \label{eq:buildC}
    C = \begin{bmatrix} 1 & 0 & 0 \end{bmatrix}
\end{equation}

We note that the internal gains represent an unobserved and biased input noise. The input and output data for all predictive controllers is normalized. The offsets are removed and it is scaled to a standard deviation of $1$. The offset removal implicitly performs an estimation and correction of the unobserved bias, under the assumption that it is constant. While this assumption is a simplification, we consider it sufficient for this application. A uniform noise with amplitude $\pm 0.05 K$, that is identical for all simulations, is added to the measurement of the zone temperature. The initialization length is $T_{ini} = 6$. This corresponds to $1 h$, which seems reasonable, considering the system's time constants. The planning horizon is $T_f = 48$, corresponding to $8 h$. Based on equation (\ref{eq:bleps}) and the variance of a uniform distribution, we get  $\varepsilon_g = 6^2 * 0.1^2/12 = 0.03$ for the bi-level DeePC. For simplicity, we use a perfect weather forecast, therefore $\tilde{\mathcal{W}} = 0$. The regularization weight for the orthogonal projection DeePC is $\lambda = 10^{5}$. The input constraints are $u \in [0, 600] W$. The cost function (\ref{eq:Jsim}) is a simple reference tracking. The optimization has no output constraints. 

\begin{equation} \label{eq:Jsim}
    J(u,y) = \lVert y - y_{Ref} \rVert_2
\end{equation}

Everything is implemented with Matlab and YALMIP. Tab. \ref{tab:solvers} lists the solvers and settings with the following abbreviations: OP for orthogonal projection, BL for bi-level and IV for instrumental variable. We use Gurobi as the default option, but switched to Matlab Quadprog for OL after Gurobi has shown poor performance for this variant. The Gurobi option BarHomogeneous refers to the use of the homogeneous barrier algorithm, which is useful for recognizing infeasibility or unboundedness, albeit a bit slower than the default algorithm \cite{BarHomogeneousGurobiOptimization}. We use it in the cases where it was suggested by an error message from Gurobi and resulted in improved performance. 

\begin{table}[h]
\caption{Solvers for all controllers}
\label{tab:solvers}
\begin{center}
\begin{tabular}{|c||l|}
\hline
Controller & Solver \\
\hline \hline
ARX-MPC & Gurobi \\
\hline
OP DeePC & Matlab Quadprog \\
\hline
BL DeePC & Gurobi, BarHomogeneous on \\
\hline
IV DeePC & Gurobi, BarHomogeneous on \\
\hline
\end{tabular}
\end{center}
\end{table}

\section{RESULTS}

This section shows the results of the building simulations in two variants, with and without internal gains, highlighting the impact of a biased input noise on the control performance.

\subsection{With internal gains}

Tab. \ref{tab:kpi} shows the RMSE of the reference tracking and the one-step lag auto-correlation of the control input. The latter is used to quantify the smoothness of the signal (high values are smoother). The significance thereof is that in many systems, jerky control inputs contribute to the degradation of the components, which may necessitate more frequent repairs. The ARX-MPC has the smallest tracking error and the smoothest control trajectory. The DeePC variants perform similarly to each other, but slightly worse than the ARX-MPC. 

\begin{table}[h]
\caption{Key performance indicators with internal gains}
\label{tab:kpi}
\begin{center}
\begin{tabular}{|c||c|c|}
\hline
Controller & RMSE [K] & Smoothness \\
\hline \hline
ARX-MPC & 0.2441 & 0.9515 \\
\hline
OP DeePC & 0.3329 & 0.9301 \\
\hline
BL DeePC & 0.3064 & 0.9346 \\
\hline
IV DeePC & 0.3328 & 0.9303 \\
\hline
\end{tabular}
\end{center}
\end{table}

\begin{figure}[thpb]
  \centering
  \includegraphics[width=3in]{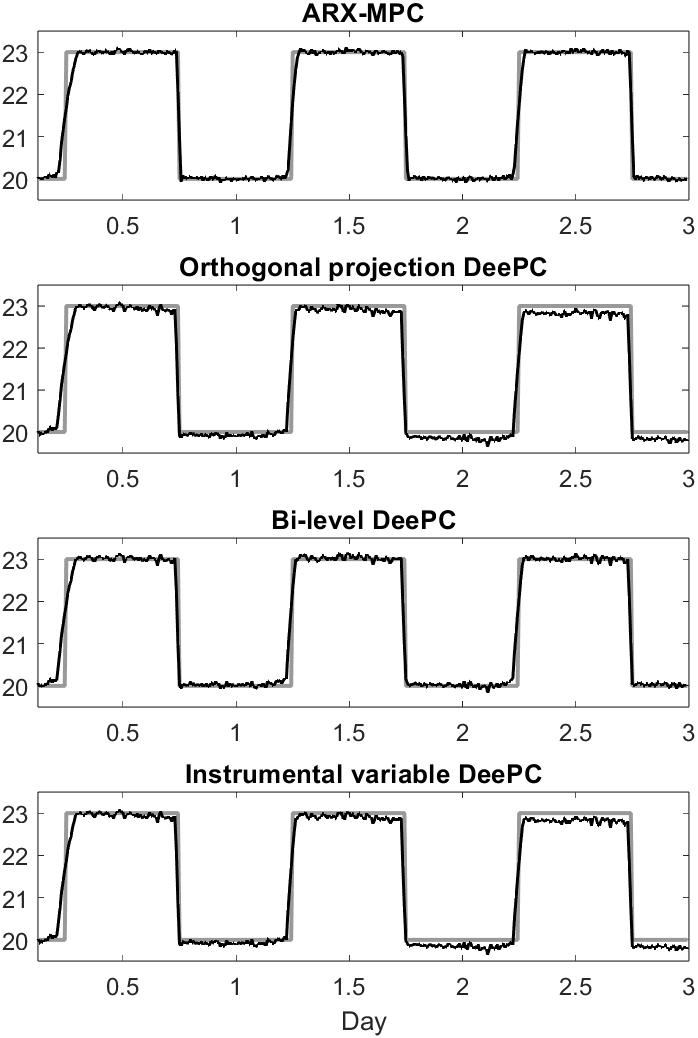}
  \caption{Reference tracking in [$^{\circ}C$] for all predictive controllers with internal gains}
  \label{fig:allY}
\end{figure}

\begin{figure}[thpb]
  \centering
  \includegraphics[width=3in]{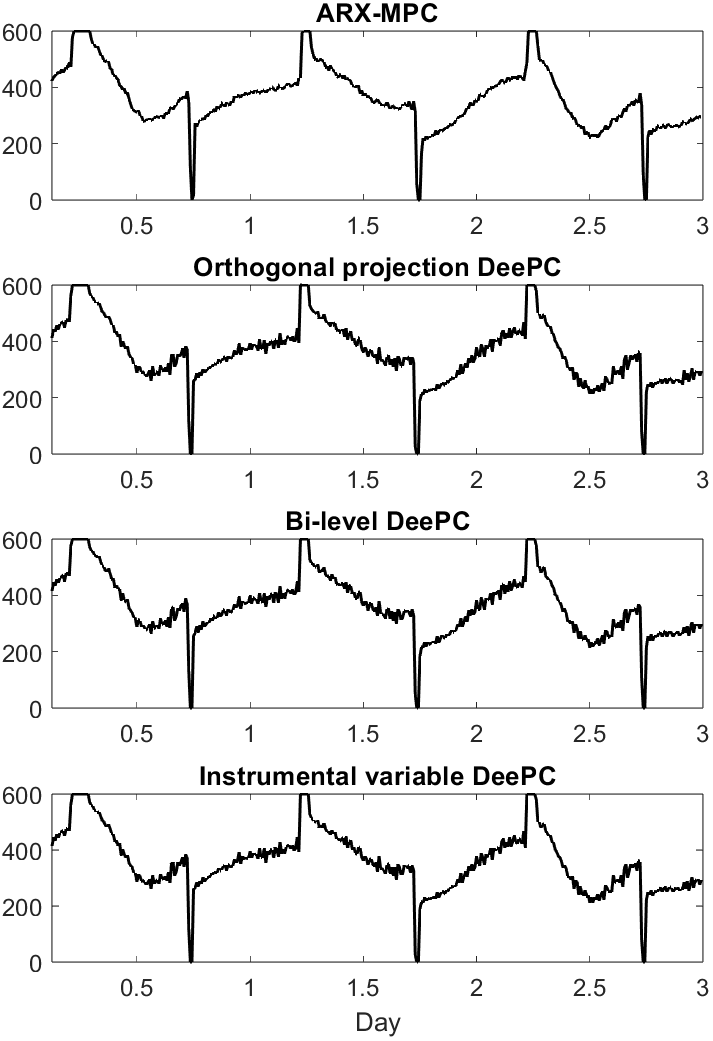}
  \caption{Control inputs in [$W$] for all predictive controllers with internal gains}
  \label{fig:allU}
\end{figure}

Fig. \ref{fig:allY} compares all system outputs (black) with the tracking references (gray). Three hours of the first day are excluded as settling time. The tracking with the ARX-MPC is accurate and smooth. The DeePC variants show visibly stronger noise in the output. The orthogonal projection and instrumental variable DeePCs also show small offsets during steady-state operation. Fig. \ref{fig:allU} shows the corresponding control trajectories, further highlighting the smoother outputs of the ARX-MPC. 

\subsection{Without internal gains}

Tab. \ref{tab:kpi_noin} shows the same performance indicators as above, but for a simulation without internal gains, i.e. the unobserved biased input noise. The values are similar to the simulation with internal gains, but the differences between ARX-MPC and DeePC are smaller. 

\begin{table}[h]
\caption{Key performance indicators without internal gains}
\label{tab:kpi_noin}
\begin{center}
\begin{tabular}{|c||c|c|}
\hline
Controller & RMSE [K] & Smoothness \\
\hline \hline
ARX-MPC & 0.2574 & 0.9507 \\
\hline
OP DeePC & 0.3174 & 0.9290 \\
\hline
BL DeePC & 0.3095 & 0.9307 \\
\hline
IV DeePC & 0.3174 & 0.9289 \\
\hline
\end{tabular}
\end{center}
\end{table}

\begin{figure}[thpb]
  \centering
  \includegraphics[width=3in]{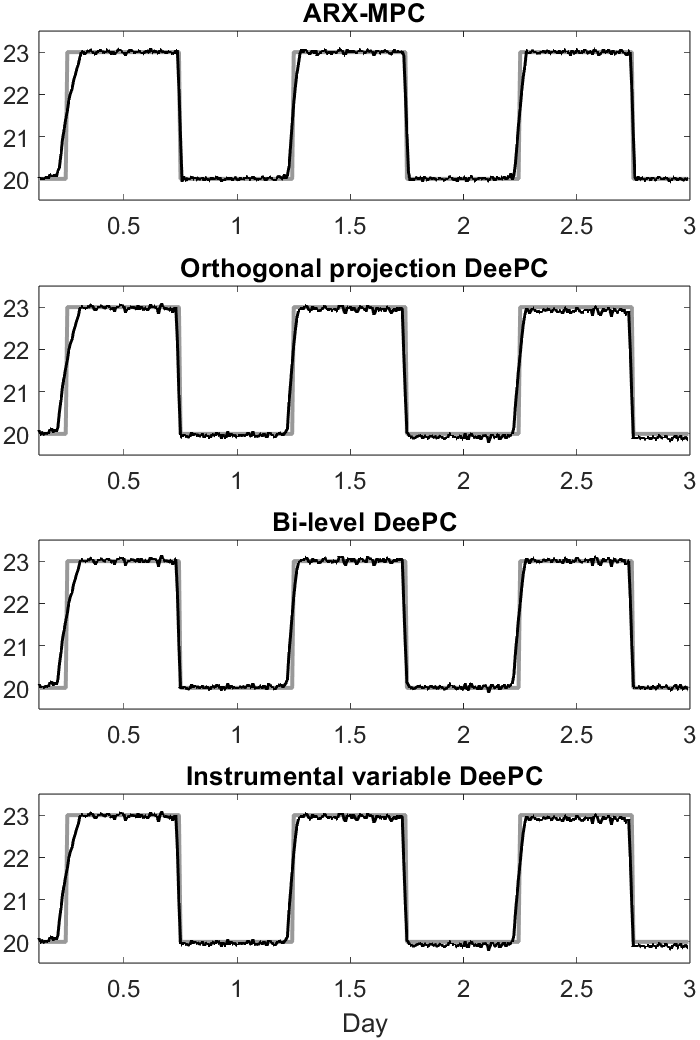}
  \caption{Reference tracking in [$^{\circ}C$] for all predictive controllers without internal gains}
  \label{fig:allYnoin}
\end{figure}

\begin{figure}[thpb]
  \centering
  \includegraphics[width=3in]{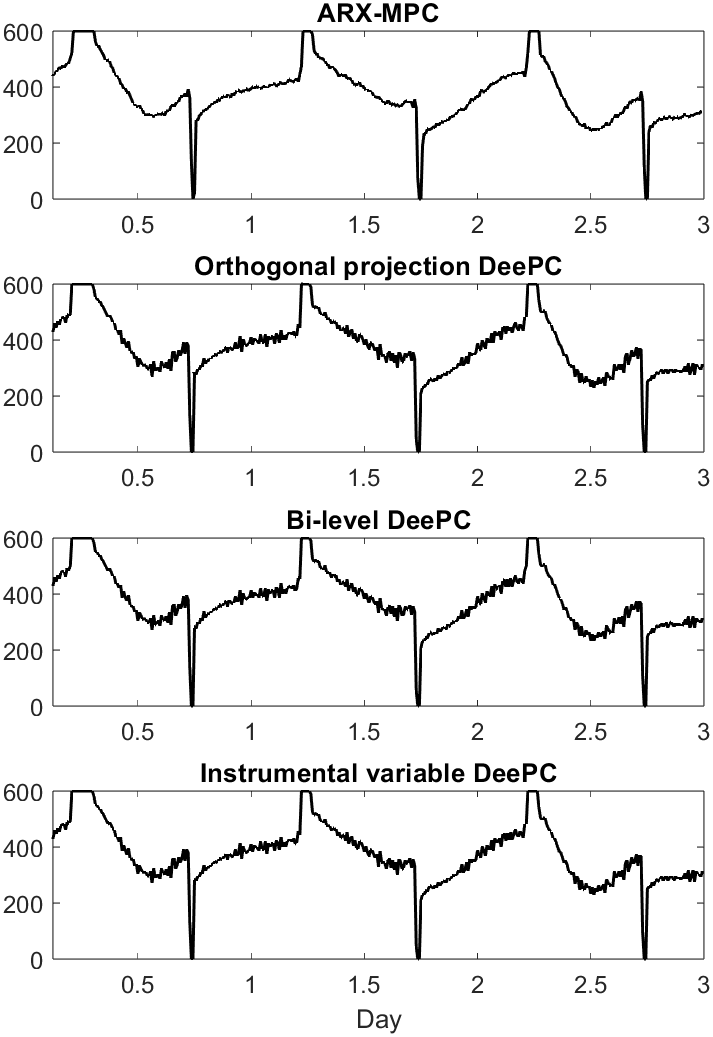}
  \caption{Control inputs in [$W$] for all predictive controllers without internal gains}
  \label{fig:allUnoin}
\end{figure}

Fig. \ref{fig:allYnoin} shows the reference tracking without internal gains. The offsets seen with the orthogonal projection and instrumental variable DeePCs above are no longer present. While offsets may be expected in the presence of unobserved biased input noise, it is noteworthy that two of the DeePC variants are affected by it, but not the bi-level DeePC or the ARX-MPC. Fig. \ref{fig:allUnoin} shows the corresponding control trajectories. No striking differences to the simulations with input noise are visible. 

\section{DISCUSSION}

Tab. \ref{tab:cmpr} compares some general qualities of the used controllers. The bi-level DeePC is by far the most complex variant in the comparison, as well as requiring a small modification of the original formulation to avoid numerical failures in our simulations. While some variants are easier to implement than others, none of them are prohibitively complex. The computational requirements on the other hand differ significantly, with two orders of magnitude between the fastest and the slowest. The bi-level DeePC differs from all others by incorporating robustness in the optimization. However, this property is not actively used in our simulation. 

\begin{table}[h]
\caption{General properties of the compared methods}
\label{tab:cmpr}
\begin{center}
\begin{tabular}{|c||c|c|c|}
\hline
Controller & Implementation & Computational & Robust \\
           & effort         & requirements  & optimization \\
\hline \hline
ARX-MPC & Medium & Low & No \\
\hline
OP DeePC & Low & Medium & No \\
\hline
BL DeePC & High & High & Yes \\
\hline
IV DeePC & Low & Low & No \\
\hline
\end{tabular}
\end{center}
\end{table}

Beyond these properties, we argue that there are two additional disadvantages of DeePC compared to the ARX-MPC from a practical point of view: The first one is the lack of quickly analyzable model properties. The poles of the ARX model (eigenvalues in state-space form) quickly reveal the stability and time constants of the plant. With the Hankel matrix of the DeePC, gaining equivalent insights requires numerical validation. The second one is that the initialization is part of an optimization that performs system identification and control optimization at the same time. As a result, the starting point of the optimized trajectory may be offset from the measured (assumed true) value. An ARX model, on the other hand, is initialized directly with measured values. The accuracy of the early steps of the optimized trajectory is usually more important than that of the steps at the end of the horizon, since only the first step is executed, before recalculation. 

We note that this study is based on one simulated system. We do not claim that our findings are complete or exhaustive, and other systems may yield different results. Furthermore, the advantages and disadvantages discussed above may be more or less relevant, depending on the field of application. 

\section{CONCLUSIONS}

Within the scope of our investigation, we find DeePC to have slightly worse performance at a higher computational cost, compared to a conventional ARX-based MPC. The three variants of DeePC all successfully eliminate the tuning process for the regularization weight $\lambda$ by incorporating a least-squares system identification in some way, which may explain the very similar performance. Furthermore, we find two of the DeePC variants to struggle with unobserved biased input noise in a way that the ARX-MPC does not. We currently have no explanation for this observation. 

\section{OUTLOOK}

Predictive control based on behavioral systems theory is an active field of research. Additional methods may be published in the near future. While this study has focused on the elimination of the regularization tuning parameters, well-developed methodologies for choosing it may solve the same problem in a different way. However, to the best of our knowledge, no such methodology is available at the time. As an example, the robust DeePC formulation in \cite{kloppeltNovelConstrainttighteningApproach2022} estimates various parameters from data, but still has two tuning weights.

\addtolength{\textheight}{-12cm}   



%




\bibliography{mybibfile}

\end{document}